\documentclass[a4paper,twoside]{article}
\newcommand{\affil}[1]{$^{\rm #1}$}
\newcommand{\HII}{H\,{\sc ii}}

\newcommand{\OIII}{[O\,{\sc iii}]}
\newcommand{\OII}{[O\,{\sc ii}]}

\newcommand{\NII}{[N\,{\sc ii}]}
\newcommand{\HeII}{He\,{\sc ii}}
\newcommand{\HeI}{He\,{\sc i}}
\newcommand{\NeIII}{[Ne\,{\sc iii}]}

\def\p0{\phantom{0}}

\usepackage[authoryear]{natbib}
\bibpunct{(}{)}{;}{a}{}{,}
\usepackage{graphicx}
\usepackage{epsfig}
\date{} 
\title{\large\bf\flushleft An Evaluation of the Excitation Class Parameter for the Central Stars of Planetary Nebulae}
\author{\parbox{\textwidth}{\flushleft
\vspace{-0.5cm}
{\it Warren A. Reid\affil{A}, Quentin A. Parker\affil{A,B}}\\
\vspace{0.4cm}
{\small \affil{A}\,Macquarie University, Sydney, Australia}\\
{\small \affil{B}\,Anglo-Australian Observatory, Sydney, Australia}\\
{\small \affil{A}\,Email: warren@science.mq.edu.au}}}
\begin{document}
\twocolumn[
\begin{changemargin}{.8cm}{.5cm}
\begin{minipage}{.9\textwidth}
\vspace{-1cm}
\maketitle
%
%
\small{\bf Abstract:
The three main methods currently in use for estimating the excitation class of planetary nebulae (PNe) central stars are compared and evaluated using 586 newly discovered and previously known PNe in the Large Magellanic Cloud (LMC). In order to achieve this we ran a series of evaluation tests using line ratios derived from de-reddened, flux calibrated spectra. Pronounced differences between the methods are exposed after comparing the distribution of objects by their derived excitation. Line ratio comparisons show that each method's input parameters have a strong effect on the estimated excitation of a central star. Diagrams were created by comparing excitation classes with H$\beta$~line fluxes. The best methods are then compared to published temperatures using the Zanstra method and assessed for their ability to reflect central star effective temperatures and evolution. As a result we call for a clarification of the term `excitation class' according to the different input parameters used. The first method, which we refer to as Ex$_{neb}$ relies purely on the ratios of certain key emission lines. The second method, which we refer to as Ex$_{\ast}$ includes modeling to create a continuous variable and, for optically thick PNe in the Magellanic Clouds, is designed to relate more closely to intrinsic stellar parameters. The third method, we refer to as Ex$_{[OIII]/H\beta}$ since the \OIII/H$\beta$ ratio is used in isolation to other temperature diagnostics. Each of these methods is shown to have serious drawbacks when used as an indicator for central star temperature. Finally, we suggest a new method (Ex$_{\rho}$) for estimating excitation class incorporating both the \OIII/H$\beta$ and the \HeII\,4686\AA/H$\beta$ ratios. Although any attempt to provide accurate central star temperatures using the excitation class derived from nebula lines will always be limited, we show that this new method provides a substantial improvement over previous methods with better agreement to temperatures derived through the Zanstra method. }


\medskip
\medskip
\end{minipage}
\end{changemargin}
]
\small

\section{Introduction}
Much progress has been achieved in modeling the central stars of planetary nebulae, seen through the progressive refinement by many authors (eg. Harm and Schwartzschild 1975; Sch\"{o}nberner 1979, 1981, 1983; Iben 1984; Wood and Faulkner 1986). Together with dynamical models for the evolution of the ejected nebula (Mathews 1966; Sofia and Hunter 1968; Kwok, Purton and Fitzgerald 1978; Kwok 1982), most authors are now modeling PN evolution with a more holistic approach (Okorokv et al. 1985; Schmidt-Voigt and K\"{o}ppen 1987a; Stasinska 1989; 2008, Dopita and Meatheringham 1990, 1992; Guenther et al. 2003; Schwarz \& Monteiro 2006; Holovatyy et al. 2008). This increasingly places more dependence on our understanding of the photoionisation of the nebula and may affect how we interpret diagnostic tools such as excitation class which has been established to give some indication of the central star's effective temperature.

Most meaningful
physical parameters that can be derived from a PN's observed emission lines, including ionized and total nebular mass
and the brightness and evolutionary state of the central star,
depend on accurate distances (Ciardullo et al. 1999). This is
difficult in our own Galaxy due to inherent problems with variable
extinction and lack of central star homoegeneity (Terzian, 1997).
The well determined 50.6~Kpc LMC distance (e.g. Keller \& Wood
2006), modest, 35~degree inclination angle and disk thickness (only
$\sim500$~pc, van der Marel \& Cioni 2001), mean that LMC PNe are
effectively co-located. Since dimming of their light by intervening
gas and dust is low and uniform (e.g. Kaler \& Jacoby, 1990), we can
better estimate absolute nebula luminosity and size. This allows us to test the various methods for deriving excitation class and place this diagnostic on a better footing.

We have constructed the
most complete, least biased and homogeneous census of a PNe
population ever compiled for a single Galaxy via discoveries from
our deep AAO/UKST H$\alpha$ multi-exposure stack of the LMC's
central 25~deg$^{2}$ (Reid \& Parker 2006a,b; Parker et al. 2005). PNe were confirmed by
2dF spectroscopy which gave 460 new PNe and independently recovered all
169 previously known PNe in the central area. 
These data have led to significant advances in our understanding of the PN luminosity function (PNLF) (Reid \& Parker, 2010) and physical
parameters such as temperatures, densities, nebula masses and
abundances (Reid 2007). Perfect for excitation studies, the LMC sample provides a large range of intrinsic flux intensities for individual emission lines (eg. 8.8E-12 to 3.7E-16 ergs cm$^{2}$ s$^{-1}$ for \OIII\,5007) and interesting differences in line ratios.


Despite all the advantages of using PNe in the Magellanic Clouds for the modeling of dynamical evolution, without deep HST imaging the central stars cannot be directly observed. In order to understand the evolution of these PNe, properties of the central star such as mass and luminosity must be modeled using properties of the nebula. Fortunately, much work has been conducted in this area and it has been convincingly argued that the strength of emission lines in the nebula are uniquely determined by the properties of the central star (Dopita et al. 1987, 1988, 1990, 1992).

Much of this central star modeling depends on some evaluation of the so-called excitation class parameter which is determined from ratios of specific emission lines, measured from the ionised nebula. In this paper, we use the large, new population of PNe in the LMC to compare the three main methods currently used for determining excitation class. Each of these different methods is somewhat grounded in the physics of low to high ionisation states but produce differing results. At last, however, with a large, more representative population of PNe in one system, covering a large evolutionary range, we can observe the different effects produced by these differing excitation schemes.

Having plotted, assessed and compared the existing excitation schemes, each is found to be reflecting different aspects of the nebula-central star relationship, none of which is fully adequate. We suggest a new method for the determination of excitation class. This method provides a smooth transition between low and high excitation regimes and allows both the helium-to-hydrogen and the oxygen-to-hydrogen ratios to contribute a weighted proportion toward the excitation estimate.

Each method is then assessed and compared by constructing diagrams, based on the standard H-R diagram. In this case, excitation class substitutes for central star temperature and the H$\beta$ luminosity of the nebula represents a rough estimate of stellar luminosity. The diagram constructed using the most correct method should provide an evolutionary map, over which previously determined evolutionary tracks should show some degree of unity if excitation class has any true intrinsic value.

\section{Spectroscopic Observations and Flux Calibration}

A major spectral confirmation program was undertaken in
November and December 2004 comprising 5 nights using 2dF on the
Anglo-Australian Telescope (AAT), 7 nights using the 1.9m at South
African Astronomical Observatory (SAAO), 3 nights using the FLAMES
multi-object spectrograph on the ESO Very Large Telescope (VLT) UT2,
7 nights using the 2.3m Australian National University (ANU)
telescope at the Siding Spring Observatory (SSO) and 3 half nights
using 6dF on the UKST. In all we obtained an unprecedented 7521 high
and low resolution object spectra for LMC targets.

Details regarding the fluxes and flux calibration are given in Reid and Parker (2010).



\subsection{Corrections for Extinction}

Extinction of light from distant objects is mainly the result of
interstellar dust. In the case of PNe, there is also extinction from circumstellar dust which dims the light of the central star and ionised nebula. Where we are using and comparing line fluxes for nebula diagnostics, it is important to correct each spectrum for both forms of dust extinction. In the optical regime, the
H$\alpha$/H$\beta$ ratio was used to determine the extinction
constant $\textit{c}$H$\beta$ (i.e., the logarithmic extinction at
H$\beta$ for each nebula. For the reddening law, the interstellar
extinction curves of Nandy et al. (1981) for the LMC were used for
the extinction functions ($\textit{f}$~($\lambda$)). 

Application of the Balmer decrement ratio 2.86 between the H$\alpha$ and H$\beta$ lines was used to correct the
spectrum in terms of the required ratios however, the value of c is dependent on flux calibration for each
of these lines and any internal inconsistencies.

The contribution of Galactic foreground dust is assumed to be low at a reddening value of $\textsl{E}$($\textsl{B--V}$)=0.074 (Caldwell and Coulson 1985). Full details may be found in Reid and Parker (2010).

\section{Determination of PN Excitation Class, \textit{E}}

There are many difficulties associated with the classification
parameters for gaseous nebulae. The first determining factor is
the temperature of the central star. The excitation class $E$ of a PN is a classification which indicates, in broad terms, the
temperature of the central ionising star as inferred from the nebula. Ratios of specific emission lines
provide a means of classifying PNe. High excitation PNe will be
those with very hot stellar nuclei. Likewise, low excitation PNe will be those with cool stellar nuclei. Although the excitation class of a
nebula relates closely to the central star, it should be viewed as a
separate parameter in its own right which may have additional
influences apart from the central star temperature or luminosity, including nebula
electron temperature, electron density, mass, chemical composition and
nebula structure. Clearly, for excitation, an independent,
quantitative method of classification is required.


The presence or absence of certain emission lines provide an initial indication of excitation. For example, the \HeII line at 4686\AA is an initial indication of high excitation.
In 49\% (207) of the newly discovered PNe, this line is not present but is commonly detected in many of the previously
known LMC PNe with only 27\% (46) non detections of \HeII\,$\lambda$4686. The presence of this
line is an initial indication of medium to high excitation of the
nebula, largely ruling out possible confusion with \HII regions.

Errors may be estimated
and applied using a combination of the line measurement errors and the flux
calibration errors. For our data this amounts to 10\% for low excitation nebulae
and 5\% for medium- and high-excitation nebulae. The errors in turn
affect the probability for any given PN to lie in a particular
excitation class, but this is not true between classes
where the presence of \HeII\,$\lambda$4686 becomes the key determinant.

As an additional test, excitation classes were determined for repeat
observations of the same PN where a different 2dF field setup was
used. A slight change in the spectrum and/or measurement of the
\OIII\,$\lambda$5007~and H$\beta$ or \HeII\,$\lambda$4686~lines would be sufficient to
shift any PN, already lying close to the edge of a class, into the
next adjoining class. Of the 88 PNe with repeat observations and determinations of
excitation class, 63 (71\%) agreed completely with the
previous class assignment. Of the remaining
PNe, 20 were away by one class, three were away by two classes (but in the
same low--medium--high bracket) and two were separated further. These
last 2 repeat observations suffered from lower S/N where the
\HeII\,$\lambda$4686~line was unable to be clearly measured. In each case the
clearest and strongest spectrum was always chosen for PN diagnostic
analysis.

\subsection{The Ex$_\mathrm{neb}$ method}

 The first
classification scheme or method we examine for determining the excitation class of a system of PNe is based on the scheme originally developed by Aller (1956) and subsequently adopted by others (eg. Feast 1968; Webster 1975; Morgan 1984; Gurzadyan 1988, 1991; Acker et al. 1992). It is arguably the most widely used excitation scheme in the literature. Using this method, the excitation classes are divided into discrete bins which may be broadly grouped into categories
described as low, middle and high. For simplicity, we refer to the two \OIII~lines, 5007\AA~and 4959\AA~as N$_{1}$ and N$_{2}$ nebulae lines.

This system is based primarily on the ratio of
$\textsl{I}$(N$_{1}$ + N$_{2}$)/$\textsl{I}$(\HeII\,4686\AA) emission
line intensities for medium and high excitation nebula and
$\textsl{I}$(N$_{1}$ + N$_{2}$)/H$\beta$ for low excitation nebula where \HeII~is no longer detected. The medium-to-high ratio is sensitive to the fraction of O$^{++}$/O$^{+}$ to be found in the nebula leading to its use in estimating the electron temperature $\textit{T}_{e}$. Where the \HeII\,4686\AA~line is extremely weak or
missing in low excitation PNe (class = 1--4) the line ratio
$\textsl{I}$(N$_{1}$ + N$_{2}$)/$\textsl{I}$(H$_{\beta}$) is used.
This ratio is linearly sensitive to the O$^{++}$/O ratio (much like
He$^{+}$/He) and exponentially to the electron temperature
$\textsl{T}$$_{e}$.
This is a very simple but effective system for excitation classification. We shall refer to this scheme as Ex$_{neb}$, because it relies solely on the measured nebula lines.

The medium and high excitation classes are denoted by integer
numbers from 4 to 12+ according to the log[$\textsl{I}$(N$_{1}$ +
N$_{2}$)/$\textsl{I}$(\HeII\,4686\AA)] ratio as shown in Table~\ref{Table
1}.

The logarithm of the result defines the high and low limits of
each class where the limits correspond to $\pm$0.1. The highest
excitation PNe are placed into the 12+ category. They have a mean
logarithmic value equal to 0.6.

\begin{table*}[]
\caption{The Ex$_\mathrm{neb}$ Excitation-Method Classification of LMC PNe}
\begin{center}
\begin{tabular}{cc|cc|cc}
  \hline
   \rule{0pt}{4ex}Low &  & Middle &  & High &  \\
  \hline
  \rule{0pt}{4ex}Class, $E$ & $I_\mathrm{N_1+N_2}/I_\mathrm{H\beta}$ & Class, $E$ & log($I_\mathrm{N_1+N_2}/I_\mathrm{H\beta}$)& Class, $E$ & log($I_\mathrm{N_1+N_2}/I_\mathrm{H\beta}$)\\
  \hline
  \rule{0pt}{4ex}1 & 0--5 & 4 & 2.6 & 9 & 1.7 \\
  2 & 5--10 & 5 & 2.5 & 10 & 1.5 \\
  3 & 10--15 & 6 & 2.3 & 11 & 1.2 \\
  4 & $>$15 & 7 & 2.1 & 12 & 0.9 \\
   &  & 8 & 1.9 & 12$^{+}$ & 0.6 \\
  \hline
  \end{tabular}
  \end{center}
\footnotesize{PNe are classified into low ($E=$1--4), middle ($E= 4$--8) and high ($E=9$--12$^{+}$)
classes. The range of line ratios is shown for each class. The basic
table layout is based on Aller (1956).}
  \label{Table 1}
\end{table*}

The medium-excitation class, $E=$~4, is considered as a transition
class where the \HeII\,$\lambda$4686~line is at the limit of detectability at
2$\sigma$ above the noise. This class may also be determined by the
high ratio $I_\mathrm{N_1+N_2}/I_\mathrm{H\beta}>15$ as shown in
Table~\ref{Table 1}. This allows PNe to be assigned an excitation class
where \HeII\,$\lambda$4686~may be below the detectability limit
but the $I_\mathrm{N_1+N_2}/I_\mathrm{H\beta}$ ratio is very large. The
value of the excitation class for any given PN is an indicator of both stellar
temperature and PN evolution, where a radius estimate can be
measured with a rate of expansion.

The fractional ionization of
oxygen rises and falls as the central star evolves. Oxygen
ionization at first rises quickly from O$^{+}$ to O$^{++}$ but then
slows as it reaches O$^{+++}$. The stellar effective temperature
($\textsl{T}_{eff}$) also increases with time as the average kinetic
energy of photoionized electrons rises from 25,000 K to over
10$^{5}$ K as the star's surface temperature also rises and it
evolves to the white dwarf phase (Dopita \& Meatheringham 1990; Stanghellini et al. 2002a).

\begin{figure}
\begin{center}
  \includegraphics[width=0.50\textwidth]{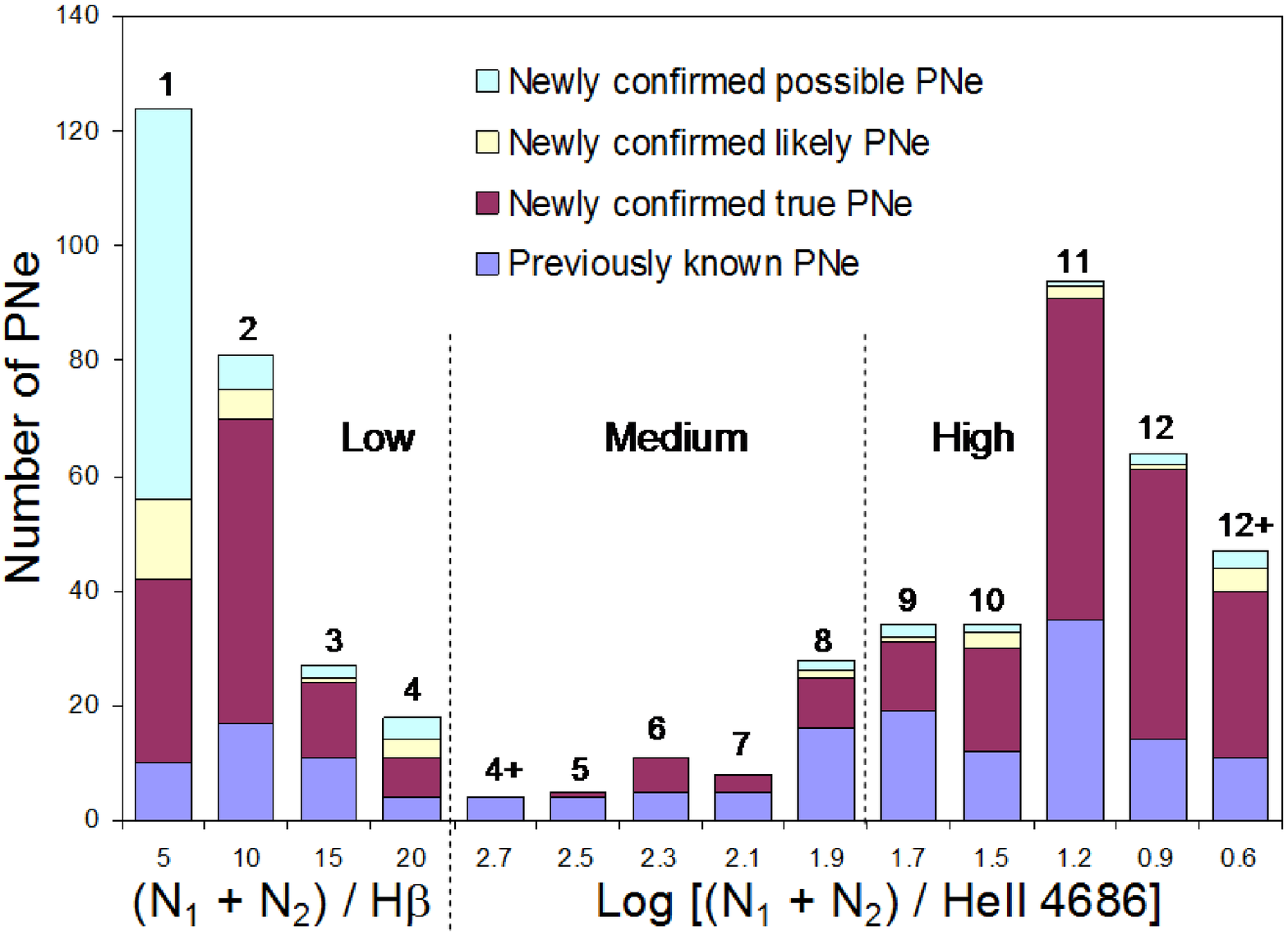}\\
  \caption{\small Graphic representation of the excitation class (Ex$_\mathrm{neb}$) for previously known and newly confirmed PNe across the whole LMC survey area. Classification is dependent upon $I_\mathrm{N_1+N_2}/I_\mathrm{H\beta}$ for low excitation PNe ($E= 1$--4) and $\log\left(I_{\mathrm{N}_{1} + \mathrm{N}_{2}}/I_\mathrm{He\textrm{\sc ii}4686}\right)$ for medium ($E=$~4--8) and high extinction PNe ($E=$~9--12$^{+}$). Newly calculated line fluxes are used throughout. The low number density of PNe in the medium excitation regime is the major feature of this method. It either points to minimal change in excitation over the life of a PN or a very fast transition between low and high regimes.}
  \label{Figure 1}
  \end{center}
\begin{center}
  \includegraphics[width=0.49\textwidth]{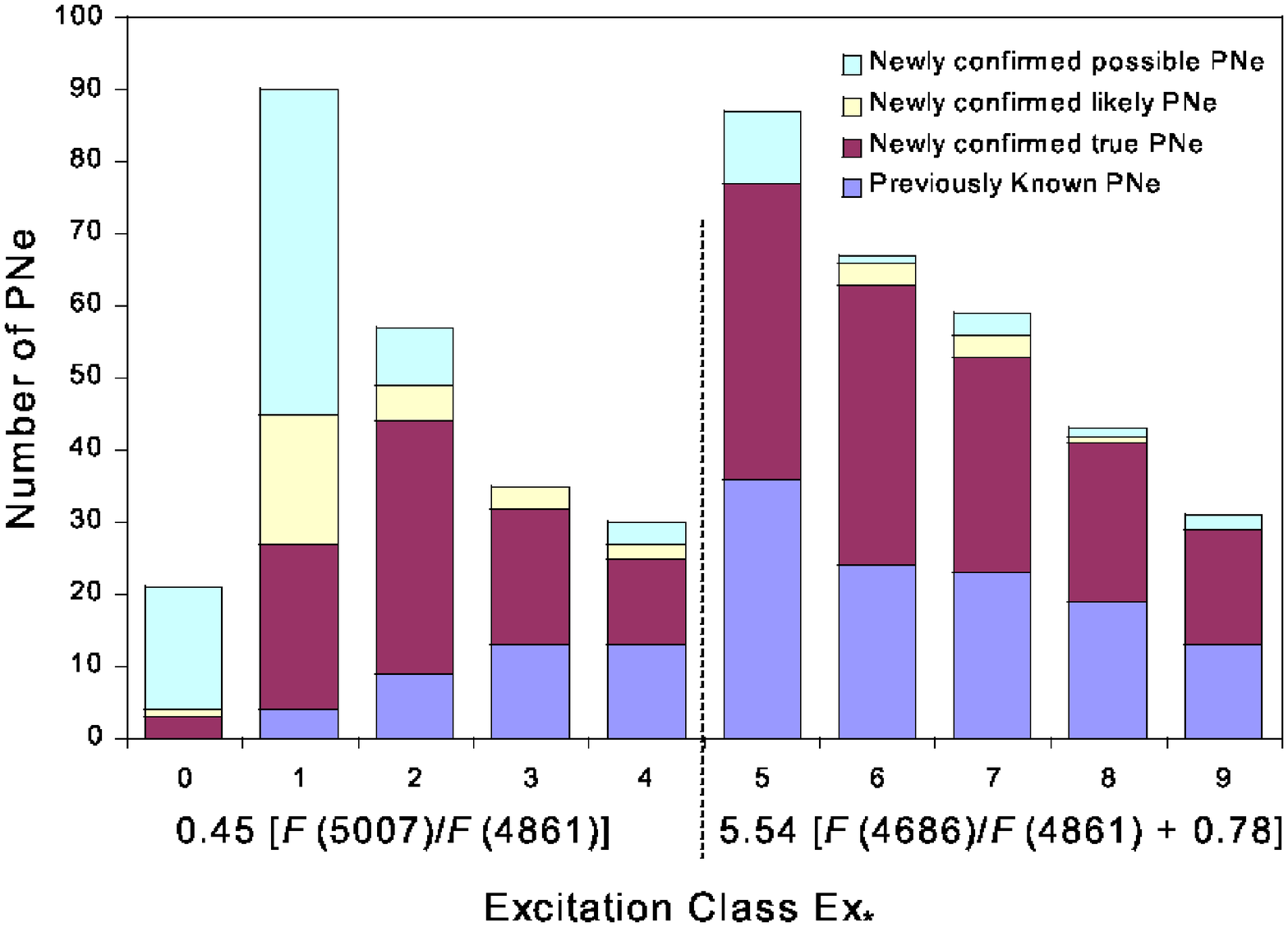}\\
  \caption{\small Graphic representation of the excitation class (Ex$_{\ast}$) based on stellar parameters for previously known and newly confirmed PNe across the whole LMC survey area. Classification is dependent upon 0.45[$\textit{F}$(5007)/$\textit{F}$(H$\beta$)] for low to medium excitation PNe (class = 0--5.0) and 5.54[0.78 + $\textit{F}$(4686)/$\textit{F}$(H$\beta$)] for medium to high excitation PNe (class = 5.0--10). Newly calculated line fluxes are used throughout. This method features a large number of PNe in the initial bin for each equation used (class 0 is a half bin). The following bins show a steady decrease.}
  \label{Figure 2}
  \end{center}
  \end{figure}
  \begin{figure}
  \begin{center}
  \includegraphics[width=0.49\textwidth]{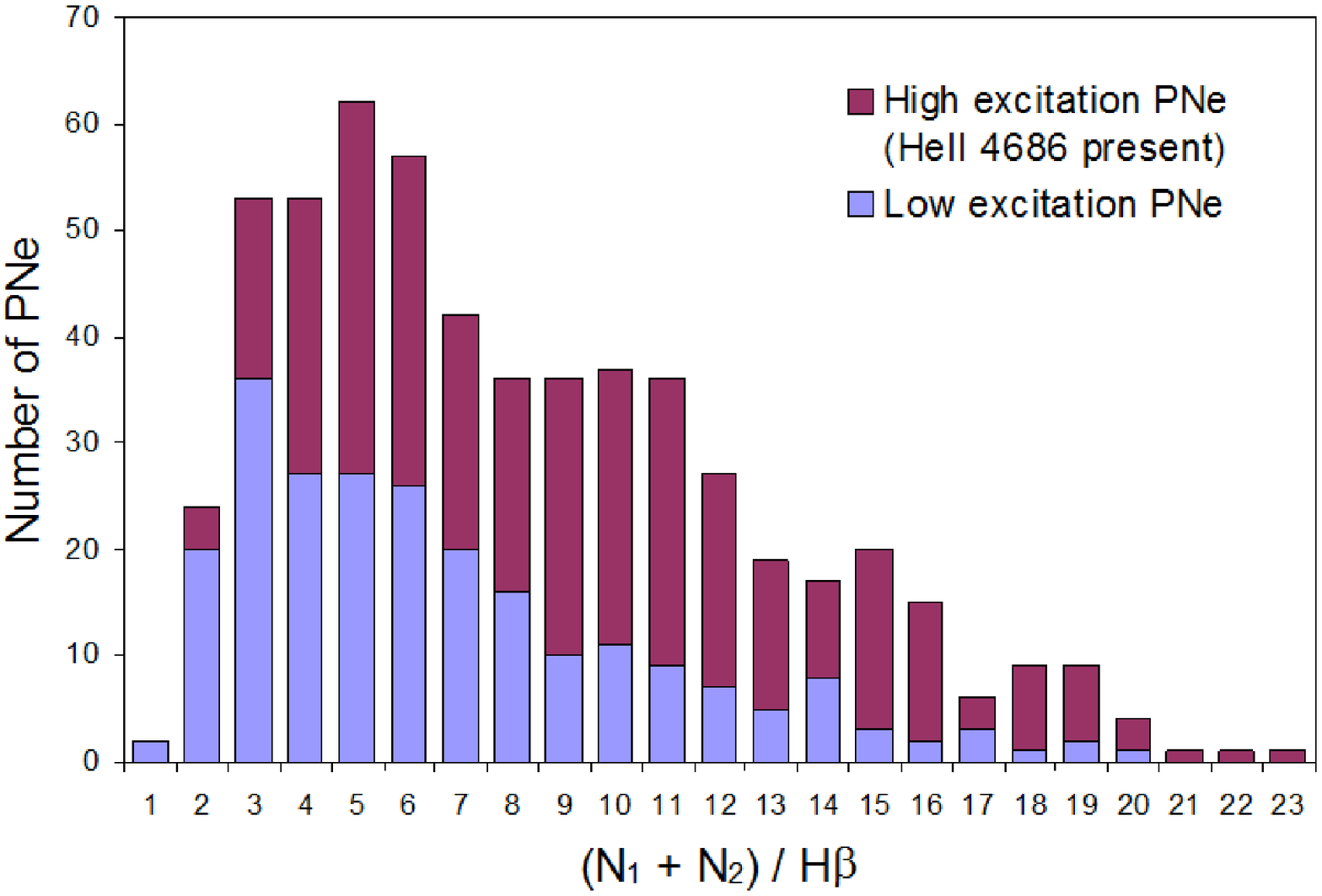}\\
  \caption{\small Graphic representation of the excitation class (Ex$_{\ast}$) based on stellar parameters for previously known and newly confirmed PNe across the whole LMC survey area. Classification is dependent upon 0.45[$\textit{F}$(5007)/$\textit{F}$(H$\beta$)] for low to medium excitation PNe ($E=$~0--5.0) and 5.54[0.78 + $\textit{F}$(4686)/$\textit{F}$(H$\beta$)] for PNe of medium to high excitation ($E=$~5.0--10). Newly calculated line fluxes are used throughout. This method features a large number of PNe in the initial bin for each equation used (class 0 is a half bin). The following bins show a steady decrease.}
  \label{Figure 3}
  \end{center}
  \end{figure}

The excitation class distribution based on this Ex$_{neb}$ scheme is presented as a histogram in
Figure~\ref{Figure 1}. 

Based on the large numbers of
new LMC PNe uncovered there are now two clear excitation classes in which most PNe may be
found during their post AGB evolution. One is at excitation classes 1-2 in the low
excitation regime and the other at classes 10-12 in the high range.
PNe in excitation classes 1-2 have spectral types earlier
than O3 on the HR diagram, which largely radiate photons with hv $<$
54.4 eV. As the excitation increases, the number of PNe decreases
towards a trough at class 5 in the medium regime before increasing
again to the high excitation peak at class 11. The low number density of PNe in the medium excitation regime is the major feature of this method. It either points to minimal change in excitation over the life of a PN or a very fast transition between low and high regimes.

Interestingly, in the high and low excitation regime, the
previously known PNe, which are generally brighter, occupy all excitation classes in
somewhat similar number ratios to the newly uncovered PNe. The exception to this is the medium excitation regime, where the brighter, previously known PNe dominate.

The strength of the combined \OIII~line intensities in most of
these PNe provides no indication of the excitation class in itself.
High excitation PNe with \HeII\,4686\AA~have central stars much hotter
than the hottest O3 stars of $\sim$46,000\,K (de Koter et al. 1998) and radiate high energy photons which
produce He$^{++}$ zones which are observed as \HeII~recombination
lines.

\subsection{The Ex$_{\ast}$ method}

An alternative approach to represent the excitation class as a continuous variable was introduced by Dopita et al. (1990; 1992)
by defining a relationship between the ratios used in the scheme of Aller (1956).
This resulted in a new classification scheme which is represented by:


 \begin{eqnarray}
 {E_\mathrm{Ex_\ast}} (0<E_\mathrm{Ex_\ast}<5.0)& = &0.45\left[\frac{F(5007)}{F(\mathrm{H\beta})}\right]\\
 {E_\mathrm{Ex_\ast}} (5.0\leq {E_\mathrm{Ex_\ast}}<10)&=&5.54\left[\frac{F(4686)}{F(\mathrm{H\beta})} + 0.78\right]~~.
 \end{eqnarray}
 
 This definition again relies heavily on the strength of the He$^{++}$ ion in classifying medium to high excitation PNe. For convenience, we shall refer to this scheme as Ex$_{\ast}$, because it attempts to create a variable to closely represent stellar temperature. There are other obvious differences in the input of the two schemes so we compare them using the large number of PNe now available in the LMC. Figure~\ref{Figure 2} shows the Ex$_{\ast}$ scheme again separated into previously known, true, likely and possible PNe and binned into discrete classes in order to compare the scheme directly to the Ex$_\mathrm{neb}$ scheme shown in Figure~\ref{Figure 1}. It is clear that Ex$_{\ast}$ moves the peak bins found in Figure~\ref{Figure 1} closer to the centre of the graph. It also has the effect of smoothing or leveling out the peaks and troughs.

 \subsection{The Ex$_{[OIII]/H\beta}$ method}

As a variation on Ex$_\mathrm{neb}$, we also measured the excitation class using $(F_\mathrm{O{\sc iii}5007}+ F_\mathrm{O{\sc iii}4959})/F_\mathrm{H\beta}$ and extended the equation across the entire luminosity range. This is the least used method but has been adopted by Stanghellini et al. (2002a) where high-excitation \HeII~lines were not available. We hereafter refer to this method as Ex$_\mathrm{[OIII]/H\beta}$. The graph for this method is shown in Figure~\ref{Figure 3} where objects containing \HeII\,$\lambda$4686 have been plotted separately. It is clear that apart from the very highest and lowest bins, this method is generally insensitive to the presence of the high-excitation \HeII\,$\lambda$4686~line. It does however provide a smooth transition across the whole evolutionary range of PNe based on the \OIII/H$\beta$ ratio alone. As an indicator of central star temperature, however, this method should not be used. Section~\ref{section 4} will show that it is undoubtedly the least useful of the three methods.

\section{Comparison of the 3 excitation methods}

Figures 1--3 clearly show that the three methods produce very different results. The differences are primarily due to the differing line ratios used for the medium- to high-excitation regimes. In all three cases, low excitation PNe are classified by \OIII/H$\beta$ in some form or other. For low excitation PNe, both Ex$_\mathrm{neb}$ and Ex$_{\ast}$ methods generally agree despite the use of slightly different line ratios. Individual PNe fall into corresponding excitation bins with a 72\% agreement. The Ex$_{\ast}$ method provides 5 low-excitation bins compared to 4 for Ex$_\mathrm{neb}$. If Bin 0 in Ex$_{\ast}$ is combined with Bin 1, a direct comparison of the 4 low-excitation bins yields an 87\% agreement. Only the Ex$_\mathrm{[OIII]/H\beta}$ method is different since it permits low excitation PNe to spread across the full bin range without restriction. Figure~\ref{Figure 3} shows that \HeII\,$\lambda$4686~may be found at low excitation levels and PNe with no \HeII\,$\lambda$4686~may be found at high excitation levels. If we take the presence of \HeII\,$\lambda$4686~as a strong indicator of medium-to-high excitation, then this method of excitation classification should be avoided.

The biggest difference between Ex$_\mathrm{neb}$ and Ex$_{\ast}$ lies in the medium to high-excitation regime. The Ex$_\mathrm{neb}$ method keeps a consistency between low and high excitation by retaining the numerator (the \OIII\,$\lambda$4959 and $\lambda$5007 line fluxes) across the whole scheme. At medium and high excitation levels the method introduces the \HeII\,$\lambda$4686~line as a denominator. This has been done in order to map a steady increase in the He$^{+}$ against O$^{++}$ species. The fact that both of these lines are likely to increase with increasing temperature is problematic. Added to this are the effects of metallicity on the strength of the \OIII~line (Dopita et al. 1992).

In the Ex$_{\ast}$ method, it is the denominator which is kept as a constant across the whole scheme. The H$\beta$ line is used as a base, against which the \HeII\,$\lambda$4686 line is allowed to increase, representing the increase in $T_\mathrm{eff}$ of the central star. The problem with this method is that H itself increases with temperature and can even be used as a broad indicator of stellar temperature (Dopita et al. 1992). The flux ratio between these two lines can be very narrow, making this discriminator extremely sensitive to line measurement errors. In addition, multiplying the result by 5.54 may have the potential to increase errors by the same factor.

\begin{figure}
  \begin{center}
  \includegraphics[width=0.48\textwidth]{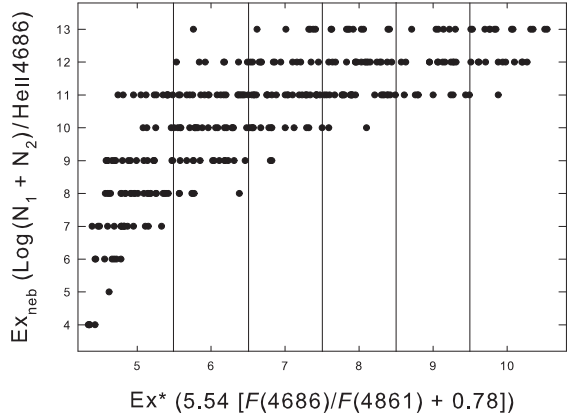}\\
  \caption{\small Comparison of excitation class $E_\mathrm{Ex_{\ast}}$ and $E_\mathrm{Ex_\mathrm{neb}}$ for the PNe of medium and high excitation. Whilst there is agreement between both methods at the highest and lowest excitation levels, the lowest excitation levels in the Ex$_{\ast}$ method include a large number of PNe estimated to be high excitation in the Ex$_\mathrm{neb}$ method. The reason for this difference is the large variation which can exist between the \OIII\,$\lambda\lambda$4959,5007 and H$\beta$ line strengths used as the basis for the Ex$_{\ast}$ and Ex$_\mathrm{neb}$ methods.}
 \label{Figure 4}
 \end{center}
  \end{figure}

Clearly, the Ex$_\mathrm{neb}$ and Ex$_{\ast}$ methods are binning PNe of medium and high excitation very differently depending on the spectral lines used, calling into question the usefulness of either scheme. In Figure~\ref{Figure 4} we compare the binning of each PNe of medium and high excitation in order to understand why the two methods produce such different results. The plot shows that $E=$~5 in Ex$_{\ast}$ contains a large number of PNe which span up to and including excitation $E=$~11 in the Ex$_\mathrm{neb}$ scheme. With each increasing Ex$_{\ast}$ excitation bin, there is an increasing agreement with the Ex$_\mathrm{neb}$ scheme. The reason for this is that we do not find high \HeII\,$\lambda$4686/H$\beta$ levels combined with low \OIII/\HeII\,$\lambda$4686~levels. This factor creates the curved, diagonal cutoff across the centre of the plot. We can therefore say that the medium excitation PNe in Ex$_\mathrm{neb}$ correspond to medium excitation in Ex$_{\ast}$. We can also say that the highest excitation PNe in Ex$_\mathrm{neb}$ agree with the highest excitation PNe in Ex$_{\ast}$. In between, however, there is some disagreement caused by a large number of PNe displaying high \OIII/\HeII\,$\lambda$4686 levels and low \HeII\,$\lambda$4686/H$\beta$ levels.

 \begin{figure}
  \begin{center}
 \includegraphics[width=0.48\textwidth]{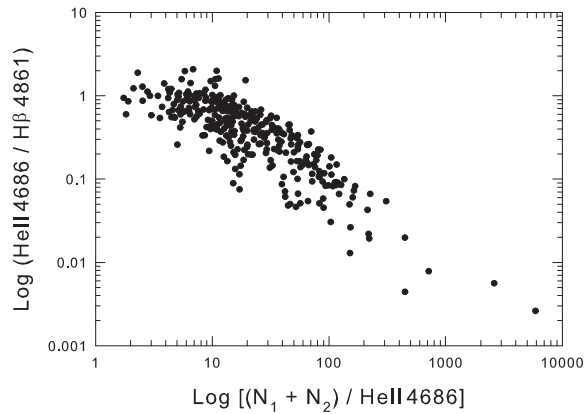}\\
 \caption{\small A comparison of \HeII\,$\lambda$4686/H$\beta$ and (N$_{1}$+N$_{2}$)/\HeII\,$\lambda$4686~for PNe of medium and high excitation. The negative correlation shows sufficient scatter to be problematic. The scatter is due to the variations in the \OIII/H$\beta$ ratios from one PN to another. The plateau on the top-left
 corresponds to cases where all the helium is doubly ionized.
}
 \label{Figure 5}
\end{center}
  \end{figure}

  \label{section 4}

  \subsection{A proposed new method for determining excitation class}

When both \HeII\,$\lambda$4686/H$\beta$ and (N$_{1}$+N$_{2}$)/\HeII\,$\lambda$4686~for PNe of medium and high excitation are directly compared, they produce a negative slope (see Figure~\ref{Figure 5}). The plateau on the top-left
of Figure~\ref{Figure 5} corresponds to cases where all the helium is doubly ionized.

\begin{figure}
  \begin{center}
 \includegraphics[width=0.48\textwidth]{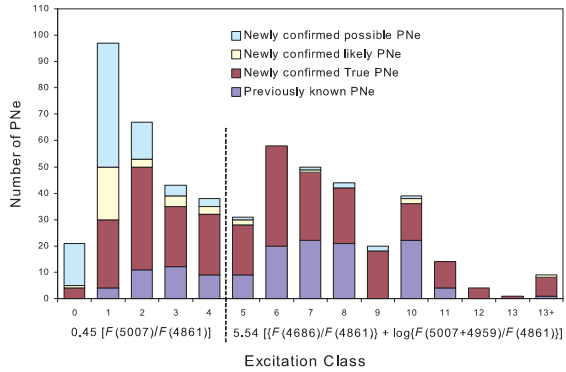}\\
 \caption{\small Graphic representation of the excitation class ($E_\mathrm{Ex_{\rho}}$) based on stellar parameters for previously known and newly confirmed PNe across the whole LMC survey area. Classification is dependent upon $0.45(F_{5007}/{F_\mathrm{H\beta}})$ for low to medium excitation PNe ($\rho$\,=\,0\,--\,5.0) and  $5.54[\log(F_{4959+5007}/F_\mathrm{H\beta})+F_{4686}/F_\mathrm{H\beta}]$   for medium to high extinction PNe ($\rho$ = 5.0\,--\,13$^{+}$). There is no longer a sudden increase in the number of PNe coinciding with the introduction of the Helium-based equation from class 5. Newly calculated line fluxes are used throughout.}
 \label{Figure 6}
\end{center}
  \end{figure}

The difference between these two methods reflects the variation between the H$\beta$\ and \OIII\,$\lambda$4959$+\lambda$5007 line strengths. 
The \OIII/H$\beta$ ratio is not at all fixed but also increases to some degree with excitation. This serves to restrain the derived excitation where \OIII/\HeII\,$\lambda$4686 levels are high, thereby providing an improved method for estimating medium and high excitation classes. The most effective way to incorporate this is to change the input parameters so that medium and high excitation become:
\begin{equation}
5.54\left[\frac{F_\mathrm{He\textrm{\sc ii}4686}}{F_{\mathrm{H}\beta}} + \log_{10}\left(\frac{F_{\mathrm{N_{1} + N_{2}}}}{ F_{\mathrm{H}\beta}}\right)\right]~~,
\end{equation}
while the low excitation PNe are estimated by
\begin{equation}
0.45\left(\frac{F_{5007}}{F_{\mathrm{H}\beta}}\right)
\end{equation}
thereby allowing the excitation class to increase in step with both the \HeII\,$\lambda$4686/H$\beta$ and the \OIII/H$\beta$ ratios. It also assigns an increasing weight to the \HeII\,$\lambda$4686~line with increasing excitation. The product values come from Dopita \& Meatheringham (1990). This results in a far more robust method for estimating excitation class and ultimately central-star temperatures. It removes the sharp cut-off between PNe of low and medium excitation, based solely on the presence of a small measure of \HeII\,$\lambda$4686, while still allowing its presence to boost the excitation level. The distribution of LMC PNe into excitation classes using this method, which we will henceforth refer to as Ex$_{\rho}$, is shown in Figure~\ref{Figure 6}.

  \section{Excitation class comparison using line fluxes}

\begin{figure}
\begin{center}
  \includegraphics[width=0.49\textwidth]{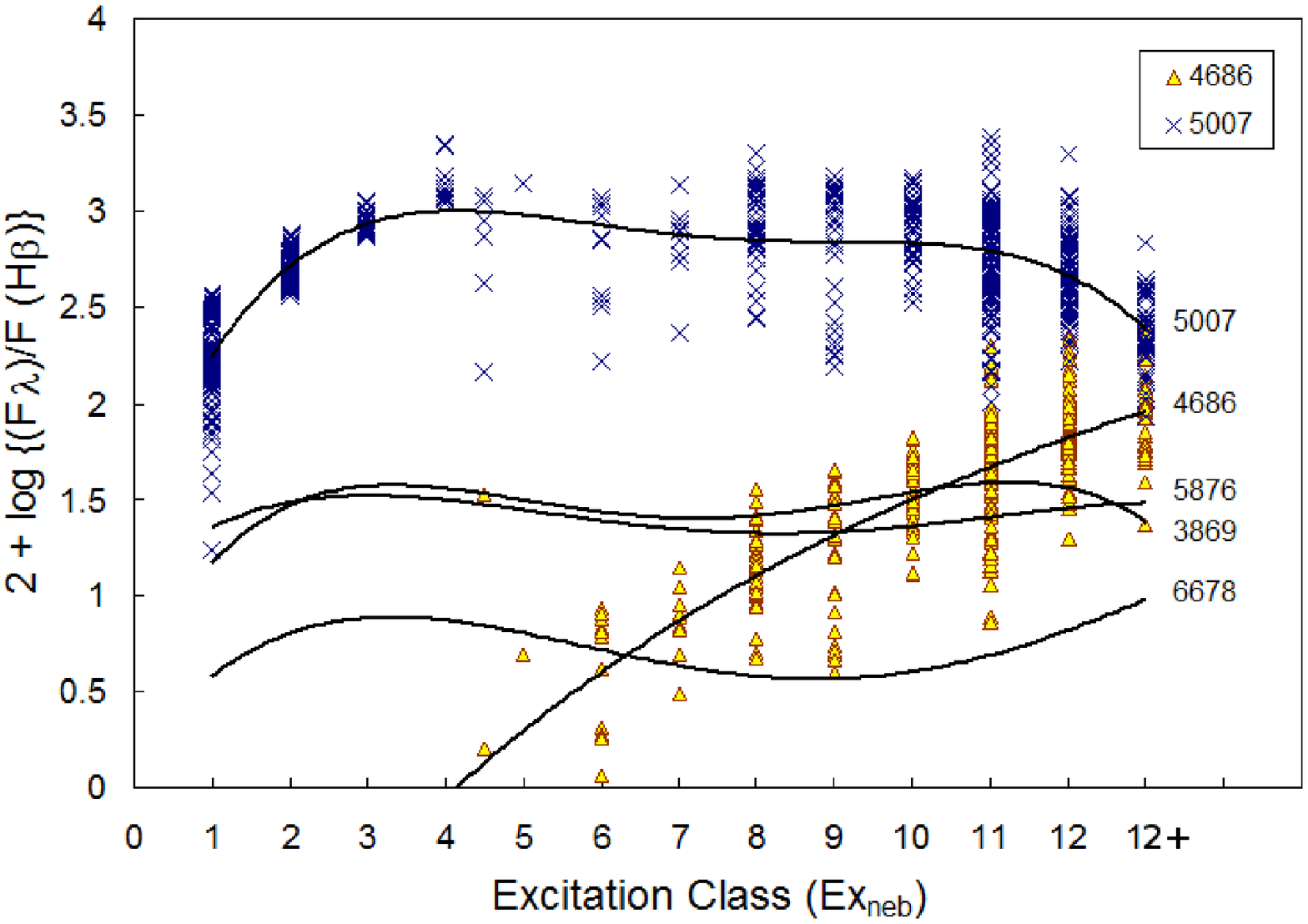}\\
  \caption{\small The position of selected high and low excitation emission line fluxes with respect to H$\beta$ for excitation classes using the Ex$_\mathrm{neb}$ classification scheme. Fluxes for \OIII\,$\lambda$5007 and \HeII\,$\lambda$4686 have been plotted together with a polynomial trend line of best fit for each line. Trend lines for \HeI\ fluxes (\HeI\,$\lambda$5876 and $\lambda$6678) and \NeIII\,$\lambda$3869 have been plotted for comparison but individual fluxes are not shown due to confusion. In this method, there is a general increase in \HeII~ flux in relation to H$\beta$ with increasing excitation. This increase is not reflected in the \OIII\,$\lambda$5007 flux.  }
  \label{Figure 7}
  \end{center}
\begin{center}
  \includegraphics[width=0.49\textwidth]{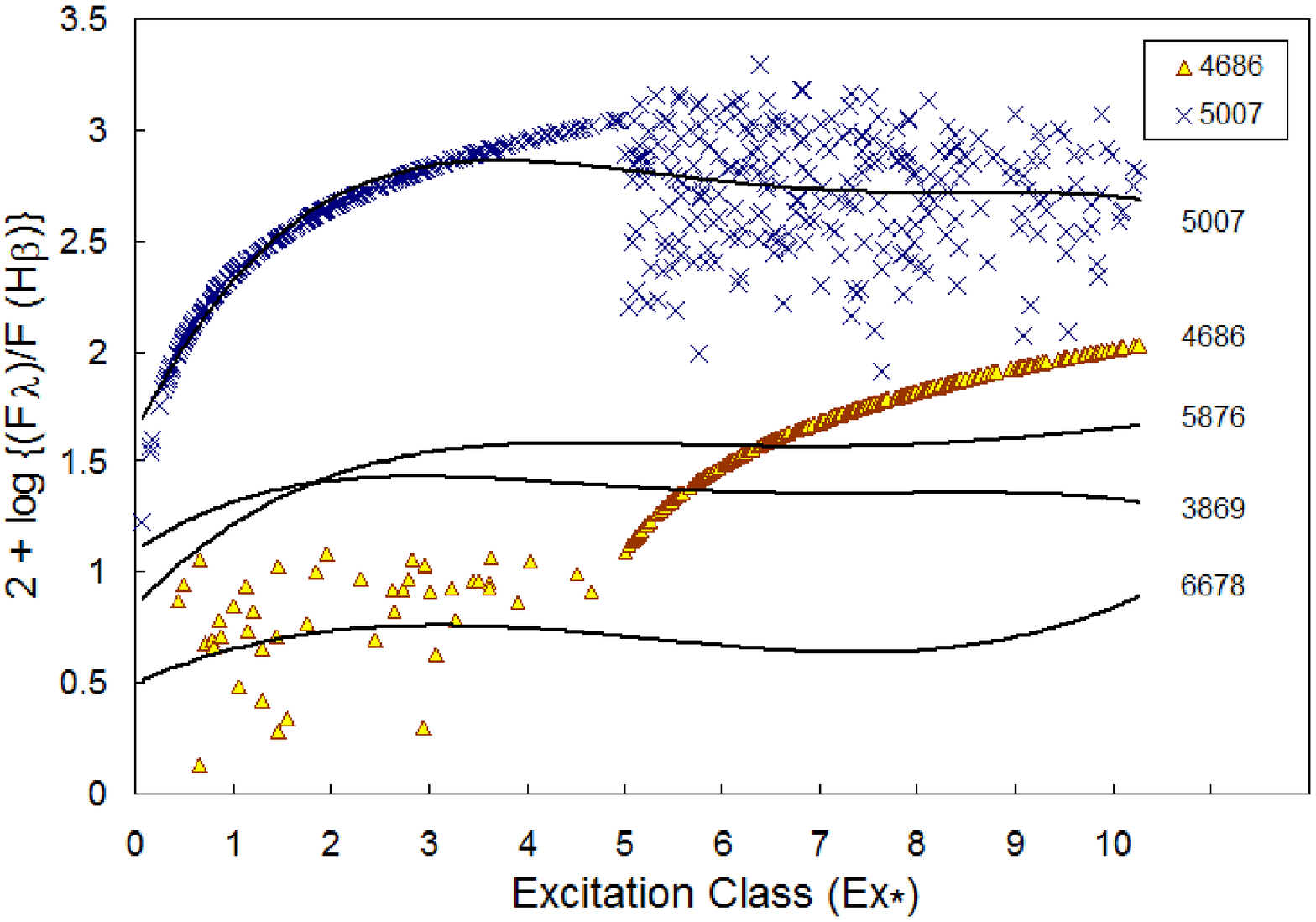}\\
  \caption{\small The position of selected emission line fluxes with respect to H$\beta$ for excitation classes using the Ex$_{\ast}$ classification scheme. Fluxes for \OIII\,$\lambda$5007 and \HeII\,$\lambda$4686~and have been plotted together with a polynomial trend line of best fit for each line. Trend lines for \HeI~fluxes (\HeI\,$\lambda$5876 and $\lambda$6678) and \NeIII\,$\lambda$3869 have been plotted for comparison but individual fluxes are not shown due to confusion. The increase in low excitation PNe is based entirely on the \OIII\,5007/H$\beta$ ratio, resulting in the exponential increase in flux with increasing excitation class. Because the medium and high excitation classes are derived from the \HeII/H$\beta$ ratio alone, the same exponential curve arises from the position this equation is applied. This has the effect of forcing too many PNe into the $E=$~5 excitation category. In a similar way, the \OIII\,$\lambda$5007 line scatters, once the low excitation equation is dropped. This method also fails to incorporate the \OIII/H$\beta$ temperature indicator.}
  \label{Figure 8}
  \end{center}
  \end{figure}

  Since each scheme uses different input parameters, we can compare the results by plotting the line fluxes in relation to H$\beta$ for each excitation scheme. This is an important comparative diagnostic because it reveals the effects of increasing excitation on the most relevant emission lines. Figure~\ref{Figure 7} shows the original Ex$_\mathrm{neb}$ scheme with \OIII\,$\lambda$5007 and \HeII\,$\lambda$4686 fluxes plotted together with a trend line. These lines are sensitive to excitation and are used in the classification scheme. The \NeIII\,$\lambda$3869 line has been used by Morgan (1984), however it was found that the scatter was rather large. We include the trend lines for the \HeI\,$\lambda$5876 and $\lambda$6678 fluxes for comparison, as a ratio of these with \HeII\,$\lambda$4686 could potentially define excitation. These fluxes were measured using the same spectra and are published in Reid (2007). The \HeI$\lambda$3889 line has been excluded as it has a metastable lower level and can therefore be strongly affected by collisional excitation and self-absorption. It is always omitted when estimating He abundances in nebulae for the same reason.

  In Figure~\ref{Figure 8} we show the \OIII\,$\lambda$5007 and \HeII\,$\lambda$4686 line fluxes as they appear using the Ex$_{\ast}$ scheme. For comparison we also include the trend line for the unseen \NeIII\,$\lambda$3869, \HeI\,$\lambda$5876, and $\lambda$6678 fluxes. The trend lines compare quite well with the Ex$_\mathrm{neb}$ method, despite individual objects being plotted in a different way. There are, however, some interesting differences which arise from the use of different equations. The \HeII\,$\lambda$4686 line, which forms a gradual curve with scatter in Ex$_\mathrm{neb}$ becomes a strict exponential in Ex$_{\ast}$. This forces a steadily increasing number of PNe into the middle excitation range. This is not surprising because the H$\beta$ line was used as the denominator in the equation. It is also used as the denominator for the $y$ axis in Figure~\ref{Figure 8}, thereby creating the continuous exponential curve. The same may be said for the \OIII\,$\lambda$5007 line in the low excitation area of Figure~\ref{Figure 8}. When the results are compared to the same graph presented as Figure~1 in Dopita \& Meatheringham (1990), the same trends are obvious but the scatter is increased in our graph due to the larger sample size and luminosity ranges.

  \begin{figure}
\begin{center}
  \includegraphics[width=0.49\textwidth]{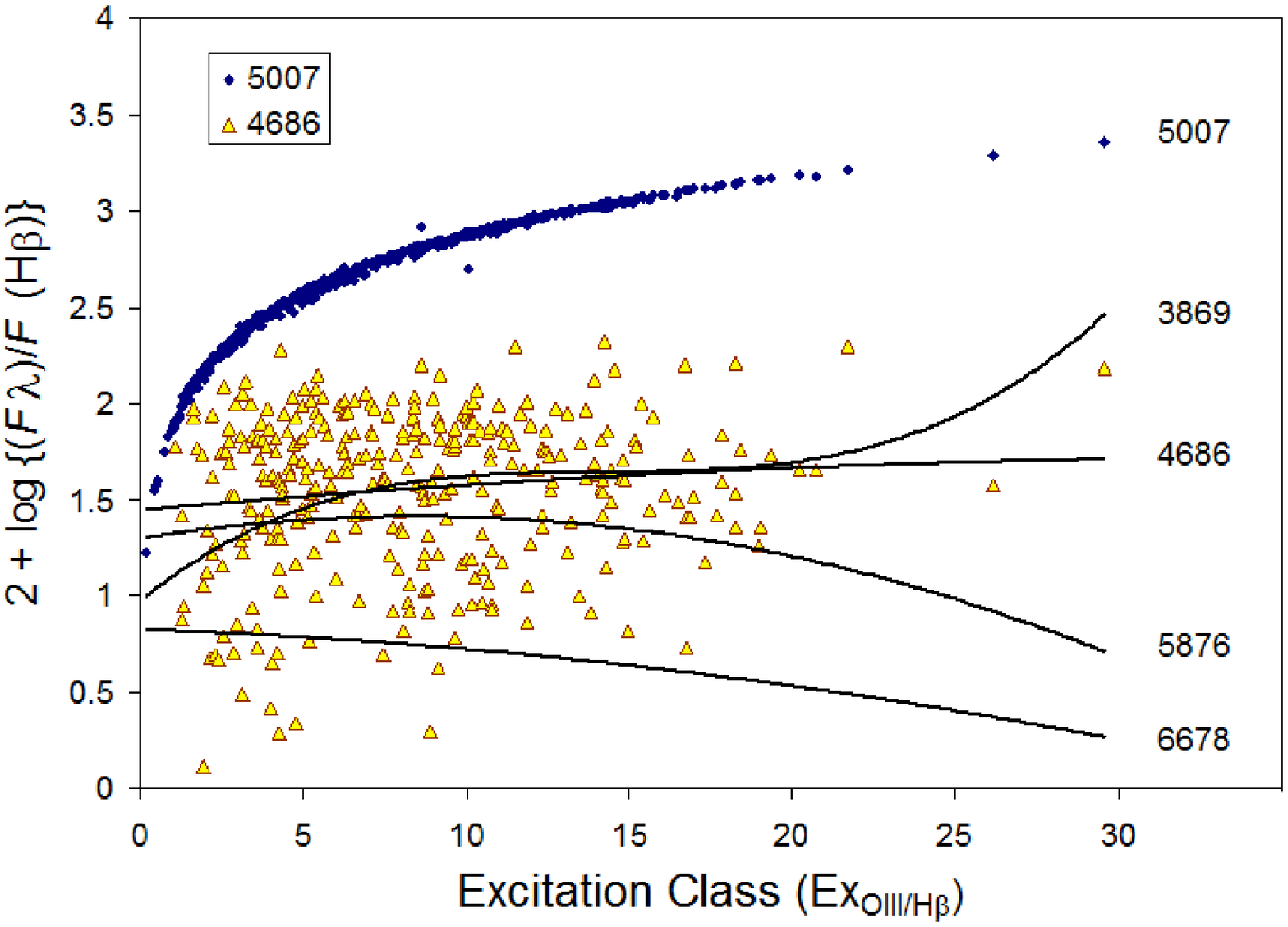}\\
  \caption{\small The position of selected emission line fluxes with respect to H$\beta$ for excitation classes using the Ex$_\mathrm{[OIII]/H\beta}$ classification scheme. Fluxes for \OIII\,$\lambda$5007~and \HeII\,$\lambda$4686~and have been plotted together with a polynomial trend line of best fit for each line. Trend lines for \HeI~fluxes (\HeI\,$\lambda$5876 and $\lambda$6678) and \NeIII\,$\lambda$3869 have been plotted for comparison but individual fluxes are not shown due to confusion. This method relies on a steady increase in the \OIII/H$\beta$ ratio across the entire excitation class range. With the \HeII\,$\lambda$4686 flux ignored in this method, it is clear that the \OIII/H$\beta$ temperature indicator provides no indication as to the \HeII\,$\lambda$4686/H$\beta$ flux ratio.}
  \label{Figure 9}
  \end{center}
  \end{figure}

\begin{figure}
\begin{center}
  \includegraphics[width=0.49\textwidth]{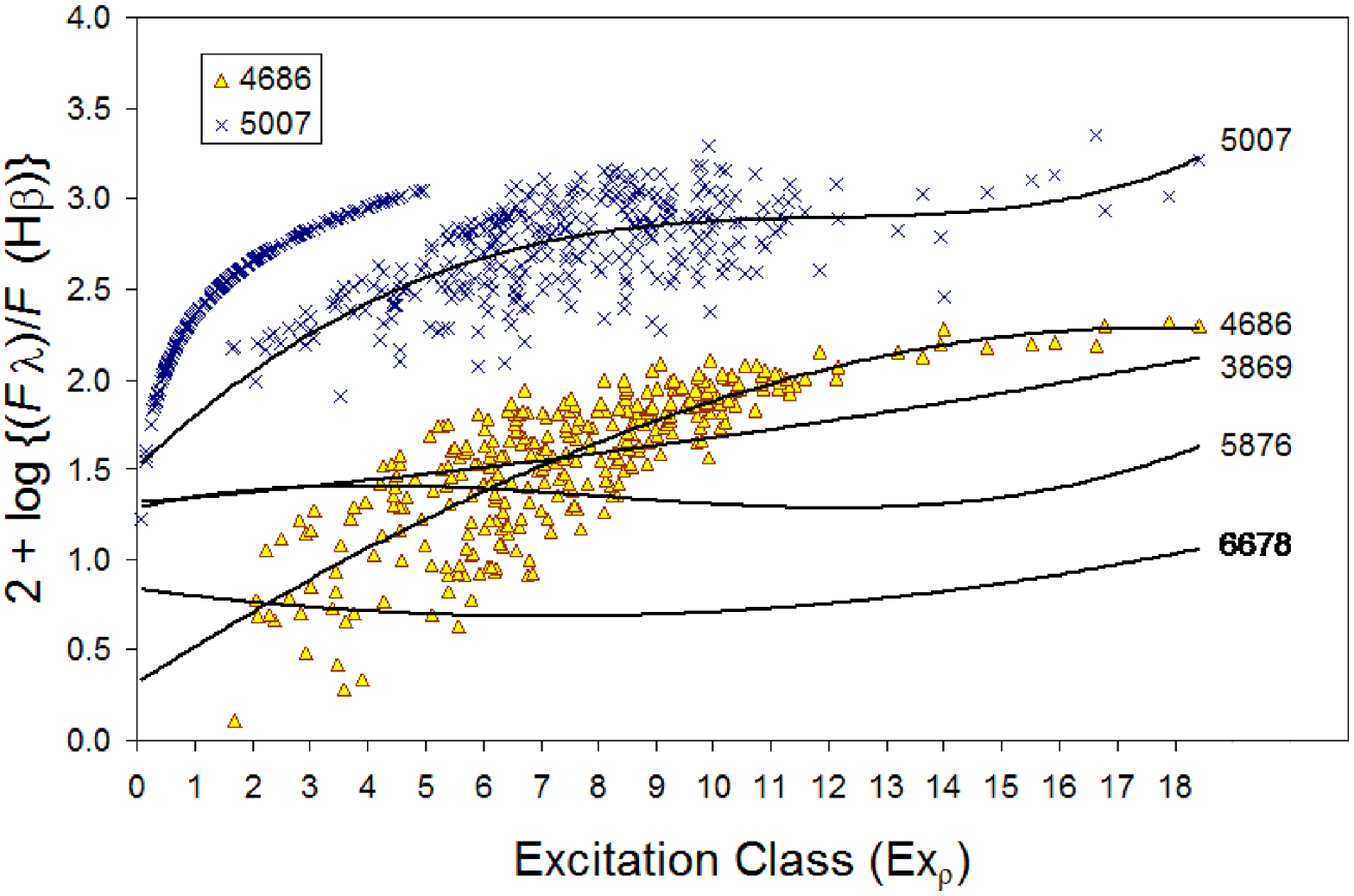}\\
  \caption{\small The position of selected emission line fluxes with respect to H$\beta$ for excitation classes using the Ex$_{\rho}$ classification scheme. Fluxes for \OIII\,$\lambda$5007~and \HeII\,$\lambda$4686~and have been plotted together with a polynomial trend line of best fit for each line. Trend lines for \HeI~fluxes (\HeI\,$\lambda$5876 and $\lambda$6678) and \NeIII\,$\lambda$3869 have been plotted for comparison but individual fluxes are not shown due to confusion. Low excitation PNe without measurable \HeII\,$\lambda$4686 occupy the classes 0 to 5 purely based on the \OIII/H$\beta$ ratio. All other PNe, which include \HeII\,$\lambda$4686 are allowed to find their excitation class position across the entire excitation range, based on both their \HeII\,$\lambda$4686/H$\beta$ and \OIII/H$\beta$ ratios. The spread at each excitation level is due to the weighted averaging between each temperature indicator. There is no sudden increase in PN at $E=$~5 where the equations meet. Both \OIII~and \HeII\,$\lambda$4686~show a steady increase with rising excitation.}
  \label{Figure 10}
  \end{center}
  \end{figure}

In Figure~\ref{Figure 9} we plot the same line fluxes using the Ex$_\mathrm{[OIII]/H\beta}$ scheme. This method relies on a steady increase in the \OIII/H$\beta$ ratio across the entire excitation class range. With the \HeII\,$\lambda$4686~flux ignored in this method, it is clear that the \OIII/H$\beta$ temperature indicator provides no indication as to the \HeII\,$\lambda$4686/H$\beta$ flux ratio. Although the minimum levels of \HeII\,$\lambda$4686~flux per excitation class somewhat increase with increasing excitation, the spread is generally even across all classes.

Finally, Figure~\ref{Figure 10} shows the same emission lines plotted using our newly suggested Ex$_{\rho}$ scheme. Low excitation PNe without measurable \HeII\,$\lambda$4686~occupy classes 0 to 5 purely based on the \OIII/H$\beta$ ratio. All other PNe, which include \HeII\,$\lambda$4686~are allowed to find their excitation class position across the entire excitation range, based on both their \HeII\,$\lambda$4686/H$\beta$ and \OIII/H$\beta$ ratios. This means that they are not restricted to begin from class 5 upwards, but may be placed in a lower excitation class depending on the line ratios.

The spread at each excitation level is due to the weighted averaging between each temperature indicator. Where \OIII~is strong and \HeII\,$\lambda$4686~is weak, more weight is apportioned to the \OIII~lines. Conversely, where \HeII\,$\lambda$4686~is strong, less weight is apportioned to the \OIII~lines. Unlike the Ex$_{\ast}$ method, on which this method is based, there is no sudden increase in the number of PNe at class 5 where the equations meet. Both \OIII~and \HeII\,$\lambda$4686~show a steady increase with rising excitation.
\section{Excitation class comparison using H$\beta$ luminosity}

If the excitation class is efficiently representing the central star temperature, then it can be used to produce a rough type of H--R diagram. For optically thick PNe, the absolute H$\beta$ flux has been suggested as a reliable indicator of stellar luminosity (Dopita et al. 1992). Diagrams using these parameters should, if the method is correct, provide a rough plot of PN evolution within a single system. Since there is a large range in CS mass and dynamical age within our LMC sample, the results appear much like scatter graphs, however, there are vital clues which may indicate which excitation methods are producing the most physically realistic results.

\begin{figure}
\begin{center}
  \includegraphics[width=0.47\textwidth]{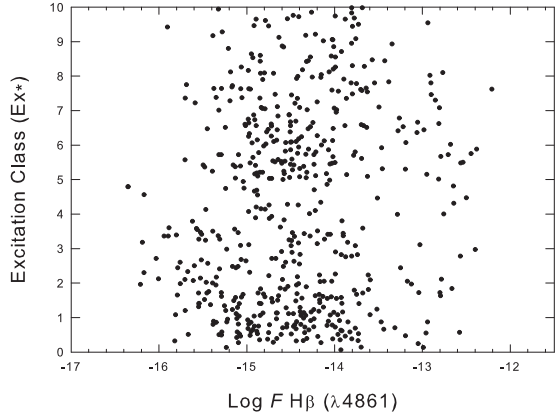}\\
  \caption{\small LMC PNe plotted according to their H$\beta$ flux and excitation class by the Ex$_{\ast}$ scheme. }
  \label{Figure 11}
  \end{center}
  \end{figure}

With the maximum luminosity conversion efficiency into the H$\beta$ line derived at a stellar temperature of near 70\,000\,K (Dopita et al. 1992), we should expect this result to be reflected in our H$\beta$ luminosities. Stellar temperatures ($t_\mathrm{eff}$) of about 70\,000\,K should be found at about excitation class 6 in the Ex$_{\ast}$ scheme and class 8 in the Ex$_\mathrm{neb}$ scheme. According to Figure~\ref{Figure 11}, this class does feature some of the brightest PNe in H$\beta$. There are a couple of outliers which somewhat obscure the result. The most notable of these is the bright PN between classes 7 and 8. This object is SMP32, a very large PN with an H$\alpha$ diameter of 12\,arcsec including the outer ionised halo (Reid \& Parker 2006b).

Our comparison confirms that most of the brightest LMC PNe do occupy the $E=$~3--6 excitation range.
PNe with lower central star temperatures are less efficient at converting luminosity into their nebulae,
making them harder to find at bright luminosity levels (Dopita et al. 1992). 


  \begin{figure}
\begin{center}
  \includegraphics[width=0.47\textwidth]{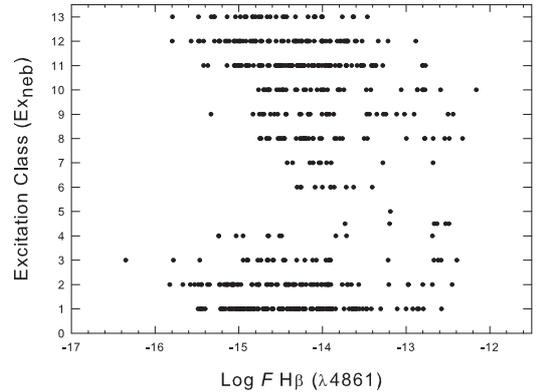}\\
  \caption{\small LMC PNe plotted according to their H$\beta$ flux and excitation class by the Ex$_\mathrm{neb}$ scheme. Medium excitation classes only have strong H$\beta$ fluxes. This excitation classification method does not allow for evolutionary motion between excitation classes due to the increasing deficit of faint PNe either side of $E=$~5.}
  \label{Figure 12}
  \end{center}
  \end{figure}

  At log$F_\mathrm{H\beta}$ fainter than --13, the number of PNe appears to increase exponentially, in agreement with earlier predictions by Henize \& Westerlund (1963) and developed by Jacoby (1989) and Ciardullo et al. (1989). Clearly, most PNe, no matter their birth luminosity, spend the largest part of their dynamical evolution $>$1\,dex below the brightest possible PN in the system. Both Ex$_\mathrm{neb}$ (shown in Figure~\ref{Figure 12}) and Ex$_{\ast}$ agree that there is a deficit of PNe in the medium excitation range (classes~4--5), now more apparent with the more representative LMC PN population available (Reid \& Parker 2006). This new finding has a huge bearing on the new LMC and MASH PN samples (Parker et al. 2006; Miszalski et al. 2008).

  The Ex$_\mathrm{neb}$ scheme in particular indicates that there are very few faint PNe in this medium excitation region. One possibility for this is that the medium excitation range is a fast, transitory period in the evolution of certain PNe. Not every PN may have the mass to enter this class but those that do will predominantly be young, optically thick PNe.

  The distribution of PNe at the bright end of both the Ex$_\mathrm{neb}$ (Figure~\ref{Figure 12}) and Ex$_{\ast}$ (Figure~\ref{Figure 11}) plots log$F_\mathrm{H\beta}>-13$ are quite similar, however, there is a considerable difference in excitation-class distribution for PNe fainter than log$F_\mathrm{H\beta}=-13.5$. The difference is due to the use of the \OIII\,$\lambda$5007 line as a denominator in the Ex$_\mathrm{neb}$ scheme. This causes the difference to be more pronounced in the mid-to-high excitation regime. Because both the input numerator and denominator for objects of medium to high excitation in the Ex$_\mathrm{neb}$ scheme are lines which increase with excitation, they push plotted objects towards high excitation levels and prevent objects being plotted where excitation and luminosity decreases. The stronger \OIII\,$\lambda$5007 in comparison to \HeII\,$\lambda$4686, the lower the derived excitation. For this reason alone, we would cast suspicion over the use of the Ex$_\mathrm{neb}$ scheme for determining excitation classes as an indication of central star $T_\mathrm{eff}$.

\begin{figure}
\begin{center}
  \includegraphics[width=0.47\textwidth]{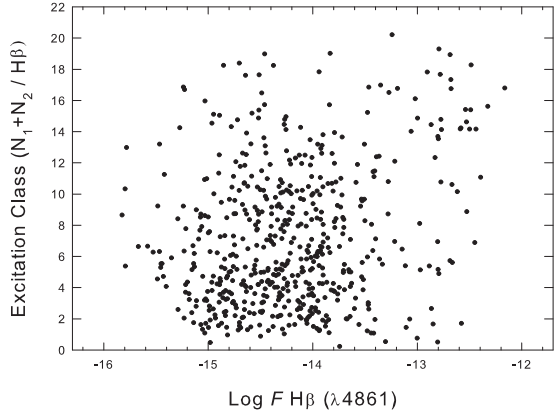}\\
  \caption{\small LMC PNe plotted according to their H$\beta$ flux and excitation class by the Ex$_\mathrm{[OIII]/H\beta}$ classification scheme. In this plot, the H$\beta$ flux is strongest between about excitation classes 15 and 18 but this is relatively equal to class 8 in Ex$_{*}$.}
  \label{Figure 13}
 \end{center}
  \end{figure}

  We also produce a diagram to compare the excitation class using the Ex$_\mathrm{[OIII]/H\beta}$ classification scheme where this equation is extended across the entire luminosity range. Figure~\ref{Figure 13} shows the result of this scheme compared to the H$\beta$\ flux. Because there is no change of input parameter at the mid-excitation range, there is a commonality between the highest and lowest plotted points. This scheme therefore provides an estimation of the shape of the excitation distribution using a single input ratio. The decreasing number of PNe found with increasing excitation is more in keeping with what we would expect to find from photoionisation modeling. There are indications, however, that this method, while internally consistent, is not producing a realistic evolutionary diagram. We would expect to find many more PNe above $\sim$\,class 17 with fluxes $\log F_\mathrm{H\beta}<-14$ since this is where we would expect to find objects of high excitation which are optically thin and therefore have much lower H$\beta$ luminosity. Most of the faintest PNe should also occupy the lowest 3--4 excitation classes.

  Finally, we produce a diagram using the new Ex$_{\rho}$ scheme (Figure~\ref{Figure 16}) developed here. The brightest PNe (brighter than $\log F_\mathrm{H\beta}=-13$) trace an upward track from class 0 to class 8 before decreasing in their H$\beta$ flux. At that point, they either move slightly downwards to join the main body of PNe or move slightly upwards, where PNe are increasingly optically thin. PNe with excitation classes greater than class 12 are extremely optically thin, making their true excitation extremely hard to plot using this method. A different excitation scheme should likely be applied to these objects. Unlike the previous excitation schemes, however, they are now appearing in the correct part of the diagram. In addition to other improvements, most of the faintest PNe are now found in the lowest four excitation classes.

  PNe may be born, evolve and dissolve at various points within these diagrams. Bearing this in mind, the Ex$_{\rho}$ scheme
  still allows us to visualise evolutionary tracks where PNe move upward at the right-hand side of the diagram, increasing in temperature, mass
  and luminosity. After reaching their brightest potential at the extreme right of the plot, they undergo a turnover where mass gain gives way to gradual mass loss accompanied by a steady decrease in luminosity and excitation class.

  \subsection{A comparison with evolutionary tracks}

\begin{figure}
\begin{center}
  \includegraphics[width=0.47\textwidth]{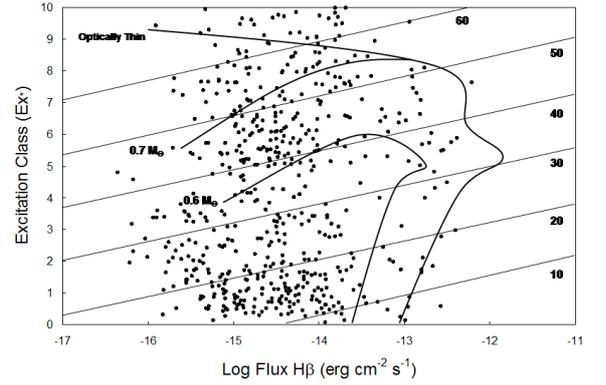}\\
  \caption{\small The Ex$_{\ast}$ method for deriving excitation class with LMC PNe plotted according to their H$\beta$ flux. The diagonal lines closely represent constant expansion velocity while evolutionary tracks show the computed cycle for 0.7 and 0.6 stellar mass central stars. From the graph, we can expect young PNe to brighten as their nebula shells accelerate. After the point of maximum luminosity, expansion begins to decline in relation to mass.}
  \label{Figure 15}
  \end{center}
  \end{figure}

   In Figure~\ref{Figure 15} we show tracks for optically thick PNe using the Ex$_{\ast}$ method. The tracks are derived in a similar fashion to that described in Dopita \& Meatheringham (1990). The tracks move from a density of 50\,000\,cm$^{-3}$ to approximately 5\,000\,cm$^{-3}$. The 0.7 M$_{\odot}$ with the highest $T_\mathrm{eff}$ has also been extended to where it would be optically thin. We include a bright extension of the tracks in class 5, which Dopita \& Meatheringham (1990) consider may exist due to atmospheric extinction of the He$^{+}$ ion in this particular temperature range. Although we have included it, we don't see any real evidence for it in the Ex$_{*}$ version of the excitation scheme. Using the Ex$_\mathrm{[OIII]/H\beta}$ method, however, shown in Figure~\ref{Figure 13}, SMP32, the very bright object sitting by itself in the other two plots may now be seen as part of an evolutionary loop or bright extension.

  \begin{figure}
\begin{center}
  \includegraphics[width=0.47\textwidth]{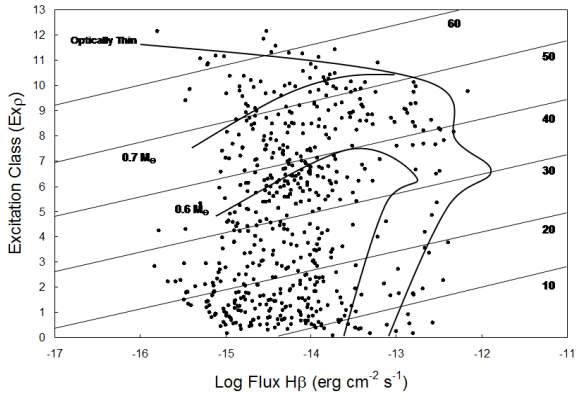}\\
  \caption{\small The Ex$_{\rho}$ method for deriving the excitation class with LMC PNe plotted according to their H$\beta$ flux. The diagonal lines closely represent constant expansion velocity while evolutionary tracks show the computed cycle for 0.7 and 0.6 stellar mass central stars. Estimated mass ranges are extended over 13 rather than 10 (see Figure~\ref{Figure 15}) excitation classes due to the different form for deriving excitation class. }
  \label{Figure 16}
  \end{center}
  \end{figure}

  In Figure~\ref{Figure 16} we show the diagram using the Ex$_{\rho}$ method in order to overlay the same evolutionary model shown in Figure~\ref{Figure 15} using the Ex$_{\ast}$ method. Estimated mass ranges are extended over 13 excitation classes rather than 10 (see Figure~\ref{Figure 15}) due to the different form for deriving excitation class. Extreme optically thin PNe, above excitation class 12 and $<$ magnitude --14, have not been plotted as their true positions cannot be correctly plotted using current methods. The positions of PNe on this plot correspond very well with the positions predicted using the evolutionary model (Dopita et al. 1990).

  \begin{figure}
\begin{center}
  \includegraphics[width=0.47\textwidth]{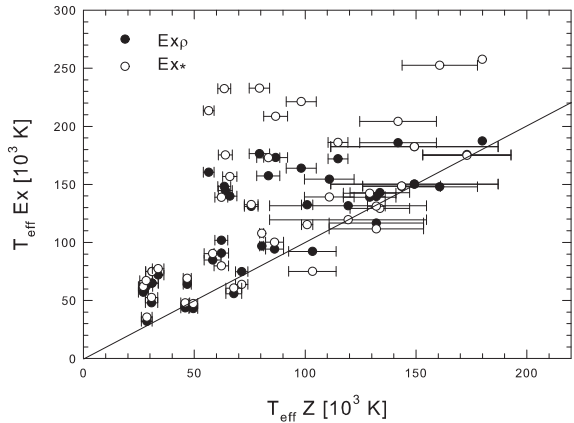}\\
  \caption{\small Temperatures derived using the Zanstra method (Villaver et al. 2003; 2007) are plotted against temperatures for the same CSPNs derived using the Ex$\rho$ and Ex$_{\ast}$ methods. A correlation coefficient of 0.742 for Ex$\rho$ and 0.616 for Ex$_{\ast}$ is found, although Ex-based methods mainly return temperatures higher by a mean of 26,000 $\pm$31,000\,K for Ex$\rho$. The line represents the 1:1 line of unity between both methods. Data for error bars are from Villaver et al. (2003; 2007).}
  \label{Figure 17}
  \end{center}
  \end{figure}

  \begin{figure}
\begin{center}
  \includegraphics[width=0.47\textwidth]{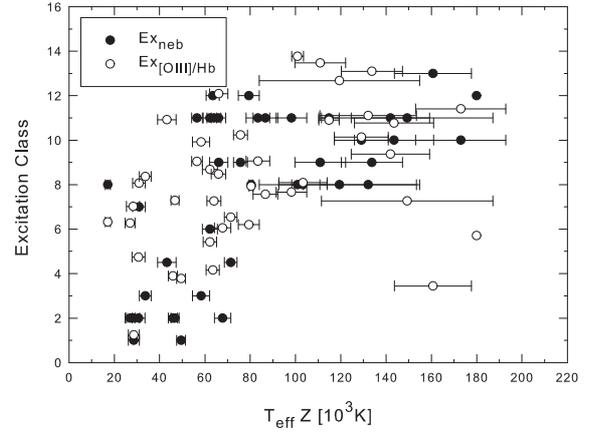}\\
  \caption{\small Temperatures derived using the Zanstra method (Villaver et al. 2003; 2007) are plotted against excitation classes for the same CSPNs derived using the Ex$_{neb}$ and Ex$_{[OIII]/H\beta}$ methods. A correlation coefficient of 0.63 for Ex$_{neb}$ and 0.37 for Ex$_{[OIII]/H\beta}$ is found. Data for error bars are from Villaver et al. (2003; 2007).}
  \label{Figure 18}
  \end{center}
  \end{figure}

  \begin{figure}
\begin{center}
  \includegraphics[width=0.47\textwidth]{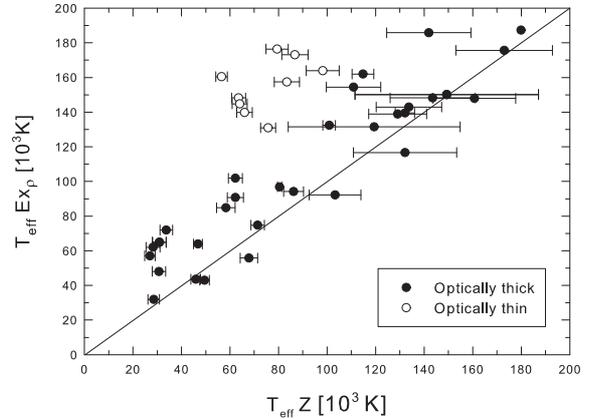}\\
  \caption{\small Temperatures derived using the Zanstra method (Villaver et al. 2003; 2007) are plotted against excitation classes for the same CSPNs derived using the newly established Ex$_{\rho}$ method. We show a separation between optically thick and optically thin PNe where the \OII3727/H$\beta$ and \NII6584/H$\alpha$ are below unity at orders less than 1.5 (Kaler \& Jacoby, 1989; Jacoby \& Kaler, 1989). In addition, T$_{eff}$Z(\HeII)/T$_{eff}$Z(H$\beta$) $\geq$ 2 for these PNe also indicates an optically thin nebula. A correlation coefficient of 0.924 for Ex$_{\rho}$ is found using optically thick PNe. The line represents the 1:1 line of unity between both methods. Data for error bars are from Villaver et al. (2003; 2007).}
  \label{Figure 19}
  \end{center}
  \end{figure}

  \section{A comparison with published temperatures}

  PN central star (CSPN) temperatures estimated through the excitation class method are compared to the most reliable CSPN temperatures for the same objects in the published literature from Villaver et al. (2003; 2007). Altogether these papers report temperature estimates for 65 CSPNs in the LMC. A total of 45 of these objects are within our survey area and are therefore used for comparison.

  To gain CSPN estimates, Villaver et al. (2003,2007) have used the method originally developed by Zanstra (1931). Although this method has been widely used (Harman \& Seaton 1966; Kaler 1983; Stanghellini et al. 2002b), the use of different nebula recombination lines produces vastly different temperature estimates. For example, the Zanstra temperature for LMC PN J33 is 38\,700 $\pm$5\,400\,K using the H$\beta$ line and $86\,700\pm$5\,400\,K using the \HeII\,$\lambda$4686 line. The difference between the two temperatures is the well-known `Zanstra discrepency' effect (Kaler 1983; Kaler \& Jacoby 1989). Our comparison tests using Zanstra temperatures calculated with the H$\beta$\ line, where the nebula is known to contain a \HeII\,$\lambda$4686~line, show scatter plots with no correlation for temperatures above 60\,000\,K. Temperatures derived using the \HeII4686 line are undoubtedly more reliable as nebulae are optically thick to \HeII\ ionising photons (Villaver et al. 2007).

  Whereas our spectroscopic data are homogenous, this is not quite true of the data used in Villaver et al. (2003,2007). Stellar photometry was obtained from HST using the Space Telescope Magnitude (STMAG) system. Extinction constants were not available in the literature for all the PNe; 6 H$\beta$ fluxes were from different authors and all \HeII\,$\lambda$5686 fluxes were gained from 5 different authors. Nonetheless, with careful treatment, the resulting temperatures probably represent the most accurate currently available in the literature to date.
  
  In Figure~\ref{Figure 17} we show a comparison of CSPN T$_{eff}$s using the Ex$\rho$ and Ex$_{\ast}$ methods against the T$_{eff}$s using the Zanstra method as published by Villaver et al. (2003; 2007). Transformation from excitation class to temperature was made using:
  \begin{equation}
  \textrm{Log} T_{eff} = 4.439 + [0.1174 (\textrm{Ex}\rho)] - [0.00172 (\textrm{Ex}\rho)^{2}]
    \end{equation}

    which is based on the transformation given in Dopita et al. (1992) but adjusted to match the published Zanstra temperatures.

  The published Zanstra-based temperature of 372\,400$\pm$97\,600\,K (\HeII\,$\lambda$4686) or 526\,000$\pm$158\,500\,K (H$\beta$) for SMP93 have been omitted as our temperature for this object is estimated at 96\,600\,K (Ex$_\rho$) or 147\,600\,K (Ex$_{\ast}$). Likewise the published, hydrogen-based $T_\mathrm{eff}$Z of 17\,100$\pm$1\,100\,K for SMP28 has been omitted since we find a strong \HeII\,$\lambda$4686 line present, raising the temperature to 81\,800\,K according to the Ex$_\rho$ result.

  Figure~\ref{Figure 17} shows a modest correlation between temperatures derived using the Zanstra method and those derived using the Ex$_\rho$ and Ex$_{\ast}$ excitation class methods. On average, Ex temperatures derived using Ex$\rho$ are 26\,000 $\pm$31\,000\,K higher than those for the same CSPNs using the Zanstra method. Much of which is due to 9 CSPNs between 50\,000 and 100\,000\,K which have $T_\mathrm{eff}$Z between 3 and 0.5 times lower than indicated by the Ex$_\rho$ method. This phenomenon has previously been observed in Zanstra-based temperature results (Kaler \& Jacoby 1989; Stasinska \& Tylenda 1986; Sch\"{o}nberner \& Tylenda 1990; Gruenwald \& Viegas 2000). It is also analogous to the test of Zanstra temperatures conducted by M\'{e}ndez, Kudritzki \& Herrero (1992) where Zanstra temperatures were compared to spectroscopic $T_\mathrm{eff}$. Low Zanstra temperatures were found for 6 CSPNs in the 60\,000K to 90\,000K range. In each case the Zanstra temperature approaches unity as the temperature increases.

In Figure~\ref{Figure 18} we show a comparison of temperatures using the Zanstra method with excitation classes determined using both the Ex$_\mathrm{neb}$ and Ex$_\mathrm{[OIII]/H\beta}$ methods. With both methods, there is a gap between PNe of low and high excitation. Low excitation PNe occupy $T_\mathrm{eff}$Z between 20\,000 and 80\,000\,K while extending up to excitation class 9. High-excitation PNe occur from class 4 upwards however occupy almost any temperature at any given excitation class. No linear fit between excitation class and temperature is possible, however, a logarithmic fit with errors up to 100\% may be possible.

Finally, we plot our new Ex$_{\rho}$ method and separate PNe which are deemed to be optically thin due to a mean \OII3727/H$\beta$ = 0.2$\pm$0.08 and \NII6584/H$\alpha$ = 0.3$\pm$0.23 with the \OII3727/H$\beta$ ratio providing the stronger indicator (Figure~\ref{Figure 19}). Plots shown by Kaler \& Jacoby (1989) and Jacoby \& Kaler (1989) indicate any N/O ratio below 0.26, corresponding to \NII6584/H$\alpha$ = 0.20 will indicate an optically thin nebula. They also allow for a factor of 5 enrichment in nitrogen by selecting N/H$\alpha$ = 1 as a secondary criterion (Kaler \& Jacoby 1989). Another indicator of optically thin PNe, $T_\mathrm{eff}$Z(\HeII)/$T_\mathrm{eff}$Z(H$\beta$) $\geq2$ (e.g. Kaler \& Jacoby 1989), is also in agreement for these PNe. A mean agreement of 83$\pm$22\% between Zanstra and Ex$_{\rho}$ temperatures and a correlation coefficient of 0.924 for Ex$_{\rho}$ are found using only optically thick PNe. This shows that, as long as optically thin PNe can be distinguished, the Ex$_{\rho}$ method will produce very good temperature estimates. Figure~\ref{Figure 19} gives us confidence that the $T_\mathrm{eff}$ of a CSPN can be determined from emission line ratios in optically thick PNe with some accuracy using the new Ex$_{\rho}$ method. Arbitrarily, the graph indicates that temperatures estimated by excitation class for optically thin PNe should be reduced by at least 50\%.

  \section{Conclusion}

The utility of the commonly used excitation class estimator for PNe is investigated and appraised using a large, new sample of PNe in the LMC.

The importance of determining the best form for deriving the excitation class is going to be an important part of our on-going study of LMC PNe. The large number of PNe now available over a wide luminosity and evolutionary range has permitted the excitation class parameter to be re-investigated and re-worked. The 3 main methods current in the literature for determining excitation class have been carefully plotted, compared and evaluated. We advise against using the Ex$_\mathrm{neb}$ scheme in its current form due to the distortion it creates at the medium-to-high excitation region. The Ex$_{\ast}$ scheme also appears to create some anomalies at the medium excitation levels where the input equation is re-worked. Using (\OIII\,$\lambda4959+\lambda5007)/\mathrm{H\beta}$ for all the PNe produces a smooth transition but suffers for not applying the \HeII\,$\lambda$4686~ratio.

The new Ex$_\rho$ method offered in this paper appears to improve excitation and stellar temperature estimates by allowing both \HeII\,$\lambda$4686 and \OIII\,$\lambda$5007\ to become constant variables against the H$\beta$ intensity. It also minimises any sudden increase in PNe at the point where the \HeII\ line is introduced into the derived equation for high-excitation PNe. The method also correlates well with the best published Zanstra temperatures, placing this new excitation class parameter on a far more solid footing.

This presentation has been prepared in order to lay the groundwork for further research using central star temperatures, electron densities, nebula mass and expansion velocities and new photoionisation codes. The disparity between the current methods for assigning excitation classes has been highlighted in this paper. Our suggested Ex$_{\rho}$ method appears to produce encouraging results as it allows excitation to increase using both the \OIII/H$\beta$ and \HeII\,$\lambda$4686/H$\beta$ CS temperature indicators.

It is clear that excitation is a quality that remains key to
the ionised nebulae well into the ageing process and must be correctly modeled. Guided by these results and comparisons, we will continue to model the excitation class based on the ratios of the same ionic species, for example the ratios of O$^{+}$ to O$^{++}$ and He$^{+}$ to He$^{++}$. This method, used already by some authors (e.g. Martin \& Viegas 2002) is expected to be a good CS temperature indicator. Our preliminary results using the LMC PNe are encouraging. With new photoionisation models, we intend refining the input parameters and anticipate that new evolutionary tracks and diagrams describing the AGB to WD stages will be produced.

\section*{Acknowledgments} 

The authors wish to thank the anonymous referee for their careful reading of the paper. We also thank David Frew for his helpful suggestions.


\end{document}